\documentclass{appolb}
\usepackage{epsfig}

\Preprinttrue
\def\bl{}
\def\blu{}
\def\mr{}
\def\dg{}

\def\scs{\scriptstyle}
\def\ss{\scriptscriptstyle}
\def\m12{m_{1\!/2}}

\def\mst{m_{\tilde\tau_R}}
\def\mchi{m_{\tilde \chi}}
\def\ohsq{\Omega_{\widetilde\chi}\, h^2}
\def\ch{\widetilde \chi}
\def\st{{\widetilde \tau}_{\scriptscriptstyle\rm R}}
\def\sm{{\widetilde \mu}_{\scriptscriptstyle\rm R}}
\def\sel{{\widetilde e}_{\scriptscriptstyle\rm R}}
\def\slp{{\widetilde \ell}_{\scriptscriptstyle\rm R}}
\def\gev{{\rm \, Ge\kern-0.125em V}}
\def\ga{\mathrel{\raise.3ex\hbox{$>$\kern-.75em\lower1ex\hbox{$\sim$}}}}
\def\la{\mathrel{\raise.3ex\hbox{$<$\kern-.75em\lower1ex\hbox{$\sim$}}}}
\def\gyr{{\rm \, G\kern-0.125em yr}}
\def\ul{\underline}
\def\tb{\tan\beta}

\begin{document}
\title{\vspace*{-2cm}\rightline{\rm MADPH-99-1144}
\rightline{November, 1999}\vspace{2cm}
Cosmological Constraints on the MSSM%
\thanks{Presented at the XXIII School of Theoretical Physics,
Ustro\~{n}'99: Recent Developments in Theory of Fundamental Interactions.}%
}
\author{Toby Falk
\address{Department of Physics, University of Wisconsin, Madison, WI~53706,
USA}
}
\maketitle
\begin{abstract}
I discuss recent developments in the study of cosmological limits on the Minimal Supersymmetric
Standard Model (MSSM).  In particular, I focus on the effect of neutralino-stau coannihilation 
on the relic abundance of neutralinos, and I give examples where the cosmologically derived limits
on the supersymmetric parameters are relaxed, and one example (CP violating phases) where they are not.
\end{abstract}
 
\section{Introduction}
The title of this talk is rather broad.  Specifically what I will be
talking about today is relic density constraints on supersymmetric models,
and I will focus in particular on bounds on minimal Supergravity
(mSUGRA).   I'll begin with a brief reminder of the notation of
supersymmetry, then give an introduction to relic densities, how they 
are computed, and how we can use them to constrain models of low-energy 
supersymmetry.  I'll then discuss coannihilation in general and show how it
dramatically relaxes the cosmological upper bound on the mSUGRA
masses.   Finally, I'll give an example of constraints (on CP
violating phases in mSUGRA)  which are {\it
  not} relaxed by the weakened mass limits.

\section{SUSY}

Since previous speakers have introduced the Supersymmetric Standard
Model, I will give only a brief reminder, in order to present the
particle content and parameters and to set the notation.  Recall that
SUSY essentially doubles the particle content of the standard
model (Table~\ref{table:susy}).  Each fermion (in fact each fermion chiral
state) has a spin-0 partner sfermion, the gauge bosons have spin-1/2 partner
gauginos, and the degrees of freedom of the Higgs sector, which now
contains two Higgs SU(2) doublets, have spin-1/2 Higgsino partners.
The four neutral gauginos and Higgsinos mix into ``neutralino'' states
$\chi_i$, so that an arbitrary neutralino is a linear combination 
\begin{equation}
  \label{eq:mix}
  \chi_i =  \beta_i  {\tilde B} + \alpha _i{\tilde W_3} + \gamma_i{\tilde
H}_1 + \delta _i{\tilde H}_2, \hspace{0.3in}  i=1,\ldots, 4
\end{equation}
where the ${\tilde B}$ and  ${\tilde W_3}$ are the partners of the
$U(1)_Y$ and neutral $SU(2)$ gauge bosons and are linear combinations
of the $\tilde\gamma$ and $\tilde Z$.
The lightest of the the neutralinos $\chi_1$ tends to be the lightest
supersymmetric particle, and in the models of interest for dark
matter, $\chi_1$ tends to be ${\tilde B}$-like, i.e. $|\beta_1|\approx 1$.
Similarly, the charged gauginos and Higgsinos mix into two ``charginos'', $\chi^{\pm}_{1,2}$.

\begin{table}[h]
\begin{center}
\begin{tabular}{rcl}
SM&&SUSY\\[0.5ex]
 \ul{fermions}&$\longleftrightarrow$& \ul{sfermions}\\[0.5ex]
${u, d, e_L,  e_R \ldots}$&&${\tilde u, \tilde d,  \tilde  e_L, \tilde e_R \ldots}$\\[0.5cm]
 \ul{gauge bosons}&$\longleftrightarrow$& \ul{gauginos}\\[0.5ex]
${g,Z,\gamma,W^\pm\ldots}$&&${\tilde g, \tilde Z,  \tilde  \gamma,\tilde W^\pm 
\ldots}$\\[0.5cm]
 \ul{Higgs bosons}&$\longleftrightarrow$& \ul{Higgsinos}\\[0.5ex]
${H_1,H_2\ldots}$&&${\tilde H_1,\tilde  H_2 \ldots}$\\[0.5cm]
\end{tabular}
\caption{\label{table:susy}The SUSY partners of Standard Model
  particles.}
\end{center}
\end{table}

Along with the new particles of the MSSM come many new (soft SUSY
breaking) parameters, including in principle separate mass parameters for all the sfermions, 
Higgs and gaugino mass parameters, and trilinear masses $A_i$ of the
Higgs-sfermion interaction terms, along
with the supersymmetric Higgs mixing mass $\mu$ and its soft SUSY breaking
counterpart $B$, and lastly the ratio of the two Higgs vacuum
expectation values, $\tan\beta\equiv v_2/v_1$:

\begin{tabular}{ll}
\\[0.1cm]
\blu Sfermion masses:&${\mr m_{\tilde f_L}^2, m_{\tilde f_R}^2}$\\[1.0ex]
\blu Higgs masses:&${\mr m_{H_1}^2, m_{H_2}^2}$\\[1.0ex]
\blu Gaugino masses:& ${\mr M_1, M_2, M_3}$\\[1.0ex]
\blu Trilinear scalar couplings:& ${\mr A_i\,\bl h_i \,\tilde Q_i \,\tilde U_i^c\, H_2 + \ldots}$\\[1.0ex]
\blu Higgs mixing masses:& ${\mr\mu\,\bl \hat H_1 \,\hat H_2}, {\mr B \mu\bl \,H_1 \,H_2}$\\[1.0ex]
\blu Higgs vev ratio \bl$v_2/v_1$\blu:& ${\mr \tan\beta}$
\end{tabular}

\vspace{0.5cm}\noindent If one includes flavor structure into the
sfermion and trilinear masses, there are over 100 new parameters
associated with softly broken supersymmetry\cite{ds}.  This large
number of parameters severely limits the predictive power of the the
MSSM, and in practice, simplifications to the set of SUSY parameters
are always made.  One of the most popular and better-motivated choices 
is inspired by minimal Supergravity (mSUGRA).  In mSUGRA, several of
the masses are taken degenerate, so that
\begin{eqnarray}\bl
m_{\tilde f_L}^2= m_{\tilde f_R}^2=m_{H_1}^2=m_{H_2}^2 & \equiv &\mr {m_{\mr0}^{\bl2}}
\bl\\
M_1=M_2=M_3& \equiv &\mr m_{1/2}\bl\\
A_e=A_d=A_u=\ldots& \equiv &\mr A_0\bl
\end{eqnarray}
at the scale $M_X$ where the gauge couplings unify.  The parameters
are then evolved to the electroweak scale using the Renormalization
Group Equations to compute the low-energy spectrum.  Due the RGE
running, the masses of the sfermions depend on both $m_0$ and
$\m12$.  The parameters $|\mu|$ and $B$ are fixed by the conditions of
correct electroweak symmetry breaking, which leaves
\begin{displaymath}
  m_0, \m12, A_0, \tb, {\rm sign}(\mu)
\end{displaymath}
as the free parameters of mSUGRA.    I will subsequently concentrate
on the mSUGRA model, but qualitatively similar results apply in the
general MSSM.

\section{The Neutralino Relic Density}
\subsection{Relic Abundances}

The possibility of a significant relic abundance of neutralinos is
partly due to R-parity, which is typically imposed on SUSY models in
order to prevent rapid proton decay. Under this new symmetry, standard
models particles have R-charge $+1$, while their superpartners have
R-charge $-1$.  Since R-parity is multiplicatively conserved, this
implies that all vertices must contain an even number of SUSY
particles, and hence that the Lightest SUSY particle (LSP) is stable.
Since it is stable over cosmological time scales, the LSP is a dark
matter candidate, and we can use limits on its relic abundance to
constrain SUSY models, as I'll describe next.

Now, the very early universe was hot and dense;  particles 
interacted rapidly,  and the LSPs, which I will denote $\chi$, were kept 
in chemical equilibrium with the standard model particles in the
thermal bath, primarily via processes in which two SUSY particles
annihilate into standard model particles, and the inverse processes in which
standard model particles annihilate to produce two SUSY particles.

\begin{figure}[h]
  \begin{center}
  \epsfig{file=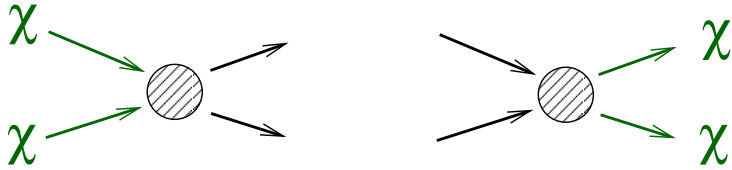, height=0.8in}     
    \caption{Keeping the $\chi$'s in chemical equilibrium.}
    \label{fig:equil}
  \end{center}
\end{figure}

At temperatures much larger than the mass of the $\chi$, the number
density of $\chi$'s was simply a spin factor times the number density of
photons.  As the universe expanded and cooled, the temperature
eventually fell below the mass of the $\chi$, and the number density of
$\chi$'s began to drop exponentially.

\begin{equation}
   {\bl n_{\dg\chi\bl}\sim\left\{
\begin{array}{rl}
{\bl{\cal O}(1) n_\gamma}& {T\gg m_{\dg\chi\bl}}\\[0.5ex]
{n_\gamma ({m_{\dg\chi\bl}\over T})^{3/2} e^{-m_{\dg\chi\bl}/T}}& {T\la m_{\dg\chi\bl}}
\end{array}
\right. }
\label{eq:neq}
\end{equation}
\noindent
If this were the end of the story, it would be a rather dull tale:
since the temperature of the universe today is about $3^\circ$
K$\sim2.5\times10^{-4}$ eV, the number density of a, say, 100 GeV $\chi$
would be suppressed vis-\`{a}-vis that of photons by a factor
$\sim\exp\{-4\times10^{14}\}$.  I.e., there would be no $\chi$'s left now.
However, in an expanding universe, this conclusion does not hold,
because at some point the $\chi$'s fall out of chemical equilibrium
with the bath.  Specifically, this occurs when the $\chi$ annihilation
rate falls below the expansion rate of the universe, $\Gamma_{\rm
  ann}\la H$.  At this point, the $\chi$'s cannot find each other
in order to annihilate sufficiently fast for their number density to
track the rapid exponential fall of (\ref{eq:neq}).  The 
number density of $\chi$'s subsequently ``freezes out'' and simply falls with the
volume of the universe, $n_\chi\sim1/V$.

\begin{figure}[h]
  \begin{center}
  \epsfig{file=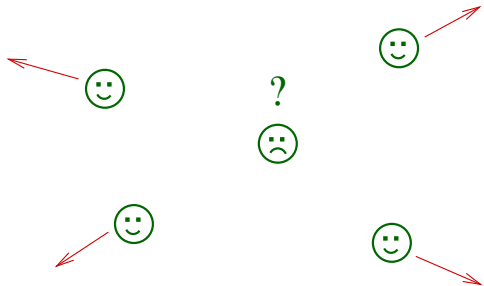, height=1.5in}     
    \caption{The $\chi$'s fall out of  chemical equilibrium.}
    \label{fig:freeze}
  \end{center}
\end{figure}

To be more explicit, the number density of $\chi$'s evolves according
to the Boltzmann equation,
\begin{equation}
  \label{eq:boltz}
  {dn_\chi\over dt}= -3 n_\chi H - \langle \sigma_{\rm ann} v\rangle
  (n_\chi^2-n_{\chi,{\rm eq.}}^2)
\end{equation}
Here $H=\dot R/R$, where $R$ is the scale factor of the universe, and
so the first term on the RHS of (\ref{eq:boltz}) simply  represents  the volume 
suppression of the number density of the $\chi$.  The first term in
parentheses describes the destruction of the $\chi$ through
annihilation, and the last term describes the production of $\chi$
particles from interactions of the thermal bath.  An approximate
analytic solution to (\ref{eq:boltz})  is given by \cite{ehnos}
\begin{equation}
  \label{eq:ohsq}
  \Omega_\chi h^2\approx {10^{-10} {\rm GeV}^{-2}\over \sqrt{\,g_f} (a +
    {1\over2}b) x_f},
\end{equation}
where $\Omega_\chi\equiv \rho_\chi/\rho_c$ is the present mass density
of $\chi$ particles in units of the critical density $\rho_c$ required 
to close the universe, $g_f$ is the number of relativistic degrees of freedom at freeze out,
$h$ is the current Hubble parameter $H$ in
units of 100 km/s/Mpc, and where the thermally averaged annihilation
cross-section has been expanded in powers of $(T/m_\chi)$:
\begin{equation}
  \label{eq:sigmav}
  \langle \sigma_{\rm ann} v\rangle = a + b \left(T\over m_\chi\right) + \ldots
\end{equation}
The temperature at freeze-out is typically well below the mass of the
$\chi$, so that $x_f\equiv T_f/m_\chi\sim 1/20 - 1/25$, and
eq.~(\ref{eq:sigmav}) is a good expansion\footnote{The temperature
  expansion is not good near s-channel resonances and just below
  important final state thresholds\cite{gs}, where the
  cross-section can vary significantly with only small variations in
  the $\chi$ energy.  However, these occur only in a limited region of
  parameter space and are not significant for us here.}.  Now, the key 
feature of  (\ref{eq:ohsq}) is that if the $\chi$ annihilation
cross-section is reduced, the $\chi$ freeze out of chemical
equilibrium earlier, when their density has had less time to track the 
exponential Boltzmann suppression (\ref{eq:neq}), and  the $\chi$ relic abundance is larger.  But a
lower bound of 12 Gyr on the age of the universe (along with the
assumption that $\Omega_{\rm tot}\le 1$) implies that  $\Omega_{\rm
  tot} h^2 \le 0.3$.  Or in other words, a lower bound on the age of
the universe implies a lower bound on the $\chi$ annihilation rate, and  this 
is the feature we will primarily exploit to constrain SUSY models.

\subsection{Relic Density Constraints}

Let us now turn to the MSSM and see what these cosmological
considerations do for us.  The lightest supersymmetric
particle is typically the lightest neutralino, and in many models,
including mSUGRA, the lightest neutralino is a quite pure bino $\tilde 
B$.  In the early universe, binos annihilate primarily via sfermion
exchange into fermion pairs.
\begin{figure}[h]
  \begin{center}
  \epsfig{file=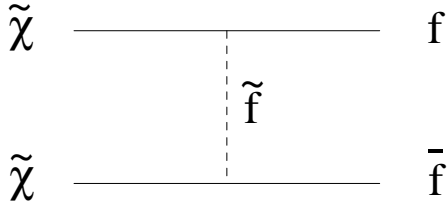, height=1.in}     
    \caption{Sfermion exchange dominates $\chi$'s annihilation.}
    \label{fig:ann}
  \end{center}
\end{figure}

Now, if the mass of the sfermions is large, then $\tilde B$
annihilation in the early universe is suppressed, and $\ohsq$ is
raised.  From the last section, we see that the lower bound on the age 
of the universe implies an {\it upper bound} on the sfermion masses, and
hence on both mSUGRA parameters $m_0$ and $\m12$.   These limits are nicely
complementary to those coming from direct searches for SUSY particles, 
which typically give lower bounds on the SUSY mass parameters.

\begin{figure}[h]
  \begin{center}
  \epsfig{file=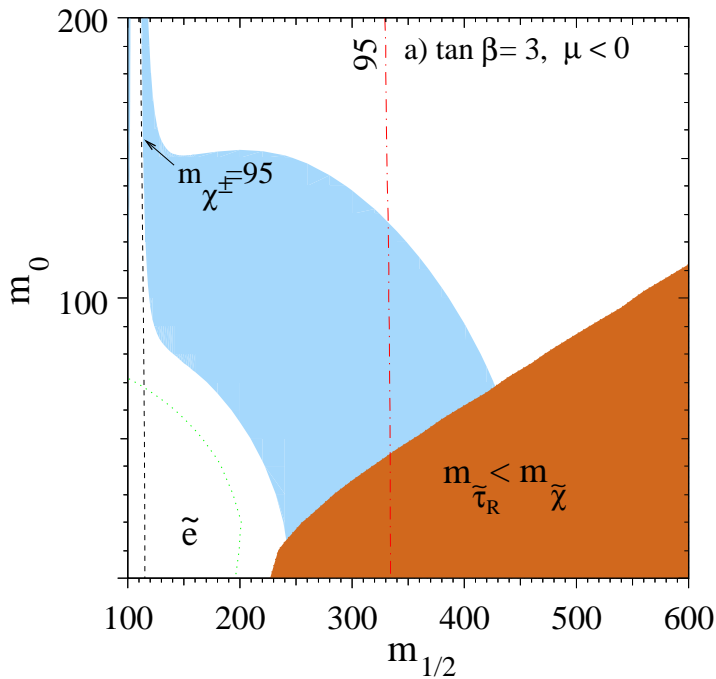, height=3.5in}
    \caption{The light-shaded area is the cosmologically preferred 
      region with \protect\mbox{$0.1\leq\ohsq\leq 0.3$}.  In the dark
      shaded regions in the bottom right of each panel, the LSP is the
      ${\tilde \tau}_R$, leading to an unacceptable abundance of
      charged dark matter.  Also shown are the isomass contours
      $m_{\chi^\pm} = 95$~GeV and $m_h = 95$~GeV, as well as an
      indication of the slepton bound from LEP.}
    \label{fig:noc}
  \end{center}
\end{figure}

The cosmological limits can be translated into $\{\m12,m_0\}$ plane \cite{efos12},
shown in Fig.~\ref{fig:noc}.  The light-shaded region corresponds to
$0.1\le\ohsq\le0.3$; the area above this region is excluded.  Below
this region, $\ohsq<0.1$, so that another component of the dark matter 
would be required.  This latter is not a bound in the same sense as
the upper limit, since we don't know for certain that any of the dark
matter is composed of neutralinos.  In the narrow chimney near
$\m12=110$GeV, $\mchi\approx m_h/2$, and s-channel annihilation
through the Higgs pole can bring the relic abundance of neutralinos
below 0.3, regardless of the sfermion masses.  In the dark shaded region, the
LSP is the right-handed stau, which is excluded by the very tight
limits on the abundance of charged dark matter \cite{ehnos}.

In Fig.~\ref{fig:noc} we also display current experimental limits:
the light dotted contour represents the bound from searches for
sleptons at LEP, while the the dashed line is a chargino isomass
contour of 95 GeV, which approximates the LEP189 chargino bounds 
at large $m_0$.  Note that the chargino bound excludes almost all of
the Higgs pole chimney.  The most significant experimental bound at this value of
$\tb$ comes from Higgs searches at LEP.  The dot-dashed contour
represents a light Higgs mass of 95 GeV, which approximates the Higgs
limit from LEP189, and the bulk of the cosmologically allowed region
is excluded.  Now, the Higgs mass itself is sensitive to $\tb$, and as 
$\tb$ is dropped, the dot-dashed contour moves quickly to the right.
It is clear that for some value of $\tb$, the Higgs contour moves to
the right of the light-shaded region entirely, and this and lower
values of $\tb$ are consequently excluded.  The current bound at this 
value of $\tb$ is around 102 GeV \cite{tamp}, and these arguments imply a lower
bound on $\tb$ of 3.7 (2.8) for $\mu <0 (\mu>0)$.  We'll see in the next
section that these constraints are weakened when we consider coannihilation.

\section{Coannihilation}

\subsection{The Basics}

So far, we have ignored interactions of the LSP with heavier SUSY
particles.  Recall that the LSPs freeze out of chemical equilibrium
when they're very cold ($\mchi/T\sim25$), so that if the mass
splitting between the LSP and the next-to-lightest supersymmetric
particle (NLSP) is ${\cal O}(1)$, the number density of NLSPs at
freeze-out is Boltzmann suppressed with respect to that of LSPs by a
factor which is $\sim\exp\{-25\}<10^{-10}$.  Therefore we don't have
to worry about NLSP interactions.  If, on the other hand, the LSP and
NLSP are closely degenerate in mass, then the NLSP interactions near
freeze-out may affect the LSP relic density.

This produces two competing effects.  First, the NLSPs freeze out of
chemical equilibrium with the standard model bath at the same time as
the LSPs and subsequently decay into LSPs, and so a significant NLSP
abundance at freeze-out can increase the relic LSP density.  Typically
a larger effect is that since the NLSP interactions contribute to the
exchange of particle number between SUSY and standard model particles
(and can dominate, as we'll see below), the SUSY particles remain in
chemical equilibrium with the thermal bath for longer and track the
equilibrium down to lower temperatures, and this reduces the LSP relic
abundance.
  
How degenerate do the LSP and NLSP states have to be in order to 
produce a significant effect?  Well, 
\begin{equation}
  \label{eq:red}
 {\frac{ n_{{\ss\rm NLSP}}}{n_{{\ss\rm  LSP}}}
\sim e^{-\Delta m/T}
\sim e^{-25(m_{\ss\rm  NLSP}/m_{\ss\rm  LSP}-1)}}.
\end{equation}
If the NLSP is 10\% (5\%) heavier than the LSP, this ratio is $\sim
{1\over10}({1\over3})$.  We see that unless the lightest states are
highly degenerate, coannihilation will only be important if
$\sigma_{\rm NLSP-LSP}$ (or $\sigma_{\rm NLSP-NLSP}$) $\gg \sigma_{\rm
  LSP-LSP}$.  And mSUGRA (and much of the MSSM), they are!!  Consider
the temperature expansion of the thermally averaged cross-section
(\ref{eq:sigmav}).  When the final state is a fermion pair (the
dominant annihilation channel for a $\tilde B$-like neutralino), $a\sim
m_f^2$.  This dependence is due to the fact that one has identical
Majorana fermions in the initial state \cite{hg} and is called
``p-wave suppression''.  Since $T/\mchi$ is small at freeze-out, this
suppresses the annihilation rate (and enhances the relic abundance) by
an order of magnitude.   Coannihilation cross-sections do not have
such a suppression and are typically an order of magnitude larger, and 
the NLSP interactions can therefore dramatically reduce the SUSY relic 
abundance.  These effects have been well studied in SUSY for
Higgsino-like neutralinos \cite{dnry}, where there is typically a close degeneracy 
between the lightest and next-to-lightest neutralinos and the lightest 
chargino.  What we have found is that coannihilation is also an essential
element in determining the cosmological upper bound on gaugino
($\tilde B$) like neutralinos, as well \cite{efosi}.

\subsection{$\tilde B-\tilde\tau$ Coannihilation}

Looking back at Fig.~\ref{fig:noc}, we shouldn't be surprised that
coannihilation may be important in mSUGRA.  The upper bound on $\m12$
occurs at the intersection of the $\ohsq=0.3$ contour with the the top
of the region with $\mst<\mchi$, i.e. at a point where the stau and
neutralino are exactly degenerate!  Generally in the MSSM$\;$\footnote{The
  presence of s-channel heavy Higgs poles can provide a loophole when
  there is a small admixture of Higgsino in the lightest neutralino
  state.}, the cosmological upper bound on the mass of the $\tilde B$
is saturated when the masses of the lightest sfermions are degenerate
with $m_{\tilde B}$.  In mSUGRA, the three right-handed sleptons
$\st,\sm$ and $\sel$ are the lightest sfermions and can all be close
in mass to the LSP.  We must therefore consider \cite{gs,efos12} the effective
annihilation cross-section
\begin{equation}
  \label{eq:sigeff}
 {\sigma_{\rm eff} = \frac{1}{n^2}\sum_{ij}\sigma_{ij} n_i^{\rm eq} n_j^{\rm eq}},
\end{equation}
where
$i,j=\st,\; \st^*, \; \sel, \; \sel^*,\sm, \; \sm^*$  and $\ch$, and
where $n=\sum n_i$.  The complete set of initial and final states
contributing to (\ref{eq:sigeff}) is given in
Table~\ref{table:states}.   The dominant contributions to $\sigma_{\rm
  eff}$ come from $\slp\slp^*$ annihilation to gauge bosons,
$\slp\slp$ annihilation to lepton pairs and $\slp\ch$ annihilation to
a lepton plus a gauge boson.  The final states with heavy Higgses turn
out to be kinematically unavailable in the regions of interest.  For 
further calculational details, see \cite{efosi}.

\begin{table}[htb]\caption{Initial and Final States for Coannihilation:
$\{i,j=\tau,e,\mu\}$}
\begin{center}
\begin{tabular}{c|l}\hline
Initial State & Final States\\ \hline\\
$\slp^{\,i}\,\slp^{\,i^{\scs *}}$    & $\gamma\gamma,\, \, ZZ\, ,\,\gamma Z,\, W^+W^-, \,Zh\, ,
\gamma h\, ,\,h \, h, \,f\bar f,$\\[0.5ex]
                & ${\rm ZH, \gamma H, ZA, W^+H^-,hH,hA, HH,HA, AA,H^+H^-}$\\[0.5ex]
$\slp^{\,i}\,\slp^{\,j}$  & $\ell^{\,i}\ell^{\,j}$\\[0.5ex]
$\slp^{\,i}\,\slp^{\,{j}^{\scs *}},\,i\ne j$   & $\ell^{\,i}\bar \ell^{\,j}$\\[0.5ex]
$\slp^{\,i}\,\ch$   & $\ell^{\,i}\gamma, \ell^{\,i}Z, \ell^{\,i}h$\\[0.5ex]
\label{table:states}
\end{tabular}
\end{center}
\end{table}

\begin{figure}[h]
  \begin{center}
    \epsfig{file=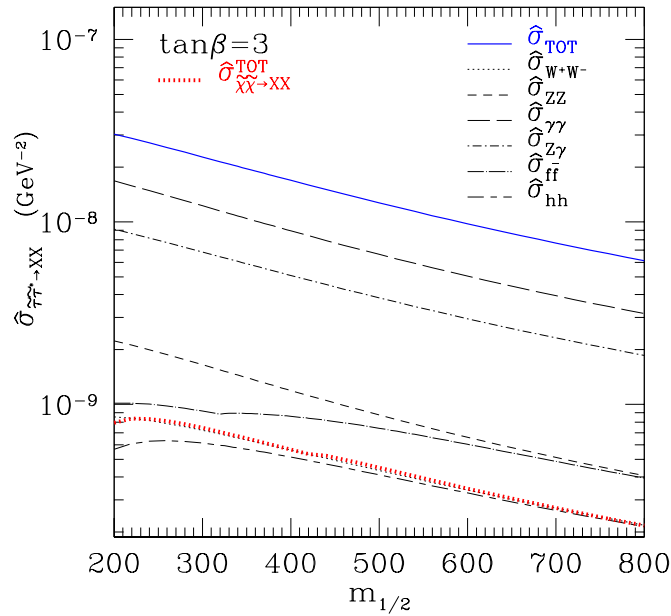, height=3.5in}
  \end{center}
\caption{\it The separate contributions to the $\st\st^*$ cross section 
    $\hat\sigma\equiv a+{1\over2}b x$ for $x=T/{m_\chi}=1/23$ and
    $m_0=120\gev$, as a function of $\m12$. For comparison, the thick
    dotted line is the $\tilde\chi \tilde\chi $ cross section. \label{fig:ss}}
\end{figure}

In Fig.~\ref{fig:ss}, we show the contributions to $\hat\sigma\equiv a
+ {1\over2}b\,x_f$ (see (\ref{eq:ohsq})) for $\st\st^*$ annihilation.
The top solid contour is the total $\hat\sigma$ for $\st\st^*$, while
for comparison we display as a thick dotted line the equivalent total
neutralino annihilation cross-section.  As advertised, the stau
cross-section is over an order of magnitude greater than that
for the neutralinos, which is p-wave suppressed.  Figures for
$\st-\tilde\chi$ and $\st-\st$ annihilation show a similar enhancement
over the $\ch-\ch$ cross-section, and figures for other $\tb$ and
$m_0$ are similar.

\begin{figure}[h]
  \begin{center}
    \epsfig{file=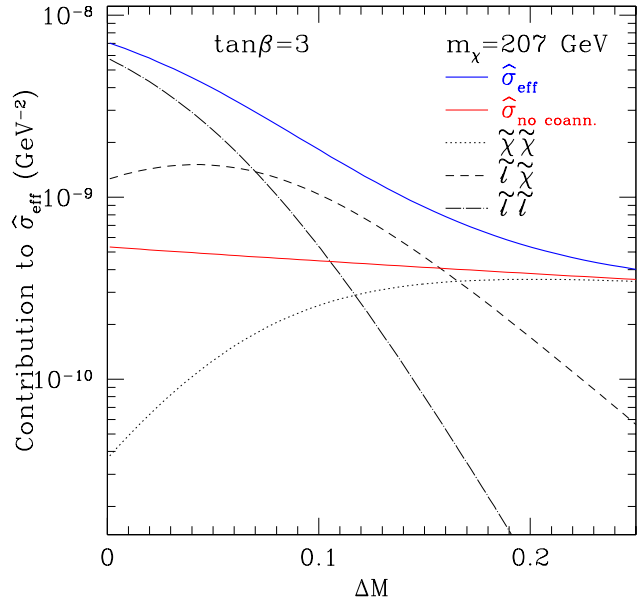, height=3.3in}
  \end{center}
\caption{{\it The separate contributions to the cross section 
    $\hat\sigma_{\rm eff}$ for $x=T/m_\chi=1/23$, as a function of
    $\Delta M\equiv(\mst- m_\chi)/m_\chi$, with $(\m12, \tan \beta) =
    (500$~GeV$, 3)$} \label{fig:svdm}}
\end{figure}

In Fig.~\ref{fig:svdm}, we display the contributions to 
$\hat\sigma_{\rm eff}$ as a function of the fractional mass difference
$\Delta M\equiv(\mst- m_\chi)/m_\chi$ between the neutralino and the
stau.  The thick solid contour shows the total $\hat\sigma_{\rm eff}$,
while for comparison the thin solid contour gives the $\hat\sigma_{\rm
  eff}$ one would compute if one ignored coannihilations, i.e.
$a_{\ch\ch} + b_{\ch\ch}/2$.  Here we've fixed $\m12=500$ GeV and
scanned upwards in $m_0$, which increases $\Delta M$.  When the neutralino and stau are
degenerate, the dominant contribution to $\hat\sigma_{\rm eff}$ comes from
slepton annihilation.  The ratio of the solid contours at this point
is greater than an order of magnitude, as above.  As $\Delta M$ increases, the density
$n_{\tilde\ell}^{\rm eq}$ of sleptons becomes Boltzmann suppressed, and the
slepton-slepton contribution falls with two powers of $n_{\tilde\ell}^{\rm eq}$
and drops below the slepton-neutralino contribution at $\Delta
M\sim0.07$.  This contribution in turn falls with one power of
$n_{\tilde\ell}^{\rm eq}$, and neutralino annihilation becomes
dominant again at $\Delta M\sim0.17$.  At large $\Delta M$, the two
solid contours and dot-dashed contour merge, and coannihilation can be 
neglected.  Again, figures for other $\tb$ are similar.

\begin{figure}[h]
  \begin{center}
    \epsfig{file=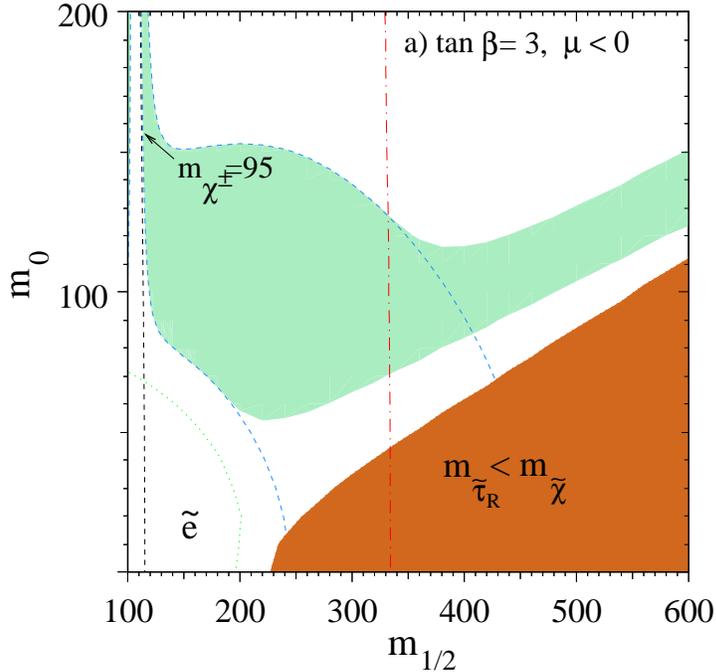, height=3.5in}  
    \caption{As in Fig.~\ref{fig:noc}, but now including
      neutralino-slepton coannihilation.}
    \label{fig:coan}
  \end{center}
\end{figure}

\begin{figure}[htb]
  \begin{center}
    \epsfig{file=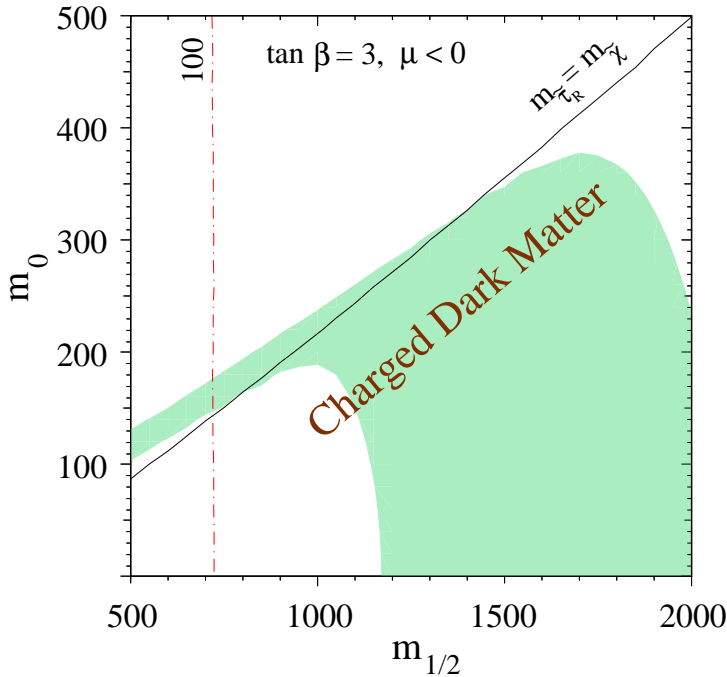, height=3.5in}  
    \caption{The same as Fig.~\ref{fig:coan}, but extended to larger $\m12$.}
    \label{fig:bg}
  \end{center}
\end{figure}

Let's now look back at the $\{\m12,m_0\}$ plane and examine the effect
of coannihilation on the cosmologically allowed region.  In
Fig.~\ref{fig:coan}, we show the same area of parameter space as in
Fig.~\ref{fig:noc}, but now with coannihilation included.  We see that
the light-shaded area now bends away from the forbidden stau LSP region and
creates a large allowed trunk which lies on top of the line
$\mst=\mchi$.  Eventually, for large enough $\m12$, the top of the
trunk falls below the $\mst=\mchi$ line, but this doesn't happen until
much larger values of $\m12$ and $m_0$, as seen in Fig.~\ref{fig:bg}.
Some features of the new cosmologically allowed region to note:
The upper bounds on $m_0$ and $\m12$ are relaxed to $\m12\la1400$ GeV
and $m_0\la350$ GeV, respectively. 
The width of the new allowed trunk is significant, from 30-50 GeV in
$m_0$ for $\m12$ up to $\sim 800$ GeV.  We've only shown the
cosmologically interesting region for one value of $\tb$,
but the shape is very similar for all small to moderate $\tb$.  The
position of the line $\mst=\mchi$ can vary somewhat with $A_0$;
however, the width of the trunk above the line is quite insensitive to 
$A_0$, as is the upper bound on $\m12$.  This relaxes dramatically the 
cosmological upper bound on the
neutralino mass in mSUGRA from about 200 GeV to close to 600 GeV.  
It will therefore take the reach of the LHC to probe the full
cosmologically interesting region.  Lastly, the bounds on $\tb$
from combining the Higgs search limits with relic density constraints
are now weakened, from 3.7 (2.8) to 2.8 (2.3) for $\mu < 0 \;(\mu > 0)$. 

\section{CP Violation and Electric Dipole Moments}

As an application, we'll now examine to what extent the relaxation of
our upper limits on $\m12$ affects constraints on CP violation in the
MSSM.  We'll start with a brief reminder of where CP violation arises
in our model.  Recall that in the MSSM, new CP violating phases
$\theta_\mu$ and $\theta_{A_i}$ accompany the (in principle) complex
parameters $\mu$ and $A_i$, introduced in the first part of this
talk.  These phases then appear in the low energy Lagrangian in the
neutralino and chargino mass matrices (in the case of $\theta_\mu$) and in
the left-right sfermion mixing terms (both $\theta_\mu$ and $\theta_A$).
The new sources for CP violation then contribute to the Electric
Dipole Moments (EDMs) of standard model fermions, and the tight
experimental constraints on the EDMs of the electron, neutron and
mercury atom place severe limits on the sizes of $\theta_\mu$ and
$\theta_A$ \cite{ko,fopr}.

The EDMs generated by $\theta_\mu$ and $\theta_A$ are sufficiently
small if either 1)~the phases are very small ($\la 10^{-2}$), or 
2)~the SUSY masses are very large (${\cal O}$~(a few TeV)), or 3)~There
are large cancellations between different contributions to the EDMs.
In mSUGRA, option 2) is forbidden by the relic density constraints, as 
we'll show next.  Condition 3), large cancellations, does naturally
occur in mSUGRA models over significant regions of parameter space, including
in the body of the cosmologically allowed region with $\m12={\cal 
  O} (100-400\;{\rm GeV})$.  These cancellations relax the constraints
on the phases, but the limit on $\theta_\mu$ remains small,
$\theta_\mu\la\pi/10$.

To see why option 2) is cosmologically forbidden, recall that the SUSY 
phases contribute to the electron EDM, for example, via processes of
the following type:

\begin{picture}(1,90)
    \epsfig{file=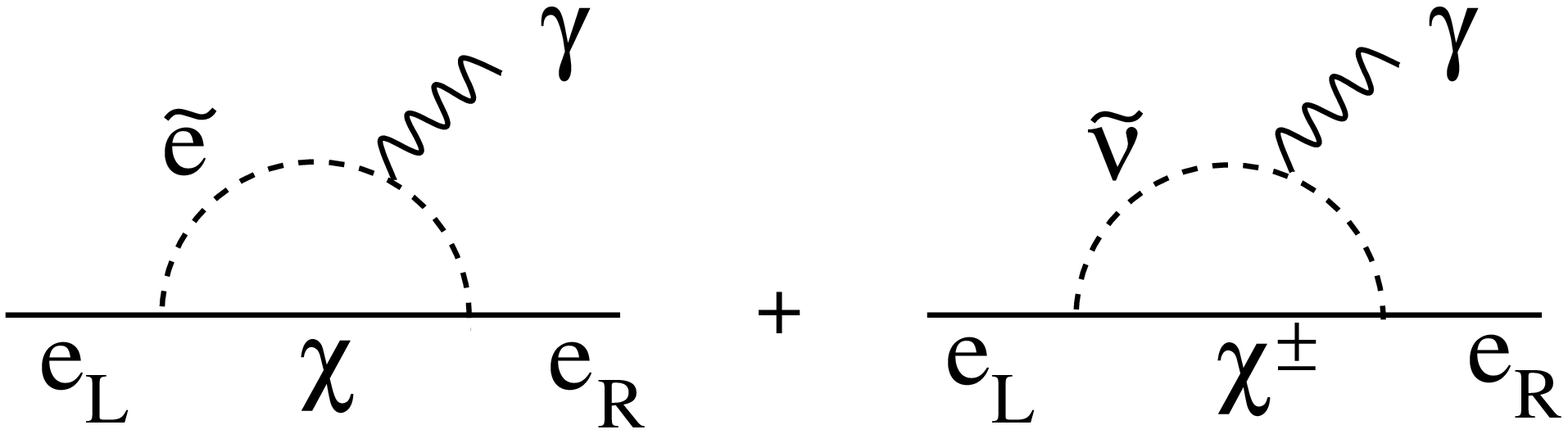, height=1.2in}     
    \label{fig:diags}
\end{picture}

\noindent{}where selectrons and sneutrinos appear in the loop.  These
contributions diminish as the sfermion masses are increased, but this
also shuts off neutralino annihilation in the early universe, which is
dominated by sfermion exchange as in Fig.~\ref{fig:ann}.  The upper
bound on $\ohsq$ then limits the extent to which one can turn off the
electron EDMs by raising the sfermion masses.  The combination of
cosmological with EDM constraints in the MSSM and mSUGRA is discussed
in detail in \cite{FOS,fopr}.

\begin{figure}[h]
  \begin{center}
    \epsfig{file=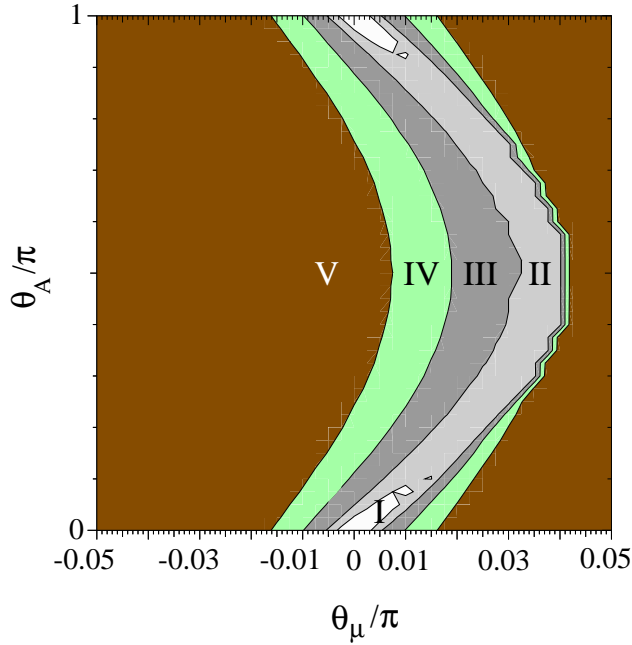, height=3.5in}
    \caption{Contours of $\m12^{\rm min}$, the minimum $\m12$ 
      required to bring both the electron and Hg EDMs below their
      respective experimental bounds, for $\tan\beta=2, m_0=130\gev$,
      and $A_0=300\gev$. The central light zone labeled ``I'' has
      $\m12^{\rm min}<200\gev$, while the zones labeled ``II'',
      ``III'', and ``IV'' correspond to \hbox{$200\gev<\m12^{\rm
          min}<300\gev$}, $300\gev<\m12^{\rm min}<450\gev$,
      $450\gev<\m12^{\rm min}<600\gev$ and $\m12^{\rm min}>600\gev$,
      respectively.  Zone V is therefore cosmologically excluded.}
    \label{fig:m12min}
  \end{center}
\end{figure}

To see the combined limits on $\theta_\mu$ and $\theta_A$ in mSUGRA,
we plot in the $\{\theta_\mu,\theta_A\}$ plane the minimum value of
$\m12$ required to bring the EDMs of both the electron and the mercury
atom ${}^{199}$Hg below their respective experimental constraints
(Fig.~\ref{fig:m12min}).  These experiments currently provide the
tightest bounds on the SUSY phases\footnote{The extraction of the
  neutron EDM from the SUSY parameter space is plagued by significant
  hadronic uncertainties \cite{fopr}, so that the inclusion of the neutron EDM
  constraint does not improve the limits when the uncertainties in the calculated
  neutron EDM are taken
  into account}.  Here we've fixed $\tb=2$,
$A_0=300$ GeV and $m_0=100$ and scanned upwards in $\m12$ until the
experimental constraints are satisfied.  Due to cancellations, the
EDMs are not monotonic in $\m12$; however, there is still a minimum
value of $\m12$ which is allowed.  Looking back at Fig.~\ref{fig:noc},
we see that in the absence of coannihilations, there is an upper bound
on $\m12$ of about 450 GeV (though slightly smaller for this $m_0$);
an analogous figure to Fig.~\ref{fig:coan} for $\tb=2$ shows that
coannihilations increase the bound to about 600 GeV.  Comparing with
Fig.~\ref{fig:m12min}, we see that zone V is cosmologically forbidden,
and that the effect of including coannihilations is to allow zone IV,
which was formerly excluded.  

Note in particular that the overall upper bound on $\theta_\mu$ in
this figure, $\theta_\mu\la0.04\pi$, is not affected by
coannihilations.  This is because the largest $\theta_\mu$ occur in
regions of cancellations, and these regions happen to lie at lower
values of $\m12$, starting in zone II with $\m12< 300$ GeV.
Increasing $m_{1/2}^{\rm max}$ from 450 to 600 GeV is insufficient to bring the
individual contributions to the EDMs to acceptable levels for the
larger values of $\theta_\mu$ and significant cancellations are still
necessary.  Even taking $\m12$ and $m_0$ at their maximal values from
Fig.~\ref{fig:coan} is not sufficient to reduce the EDMs below their
experimental limits, and so coannihilation {\it does not affect} the
upper bound on $\theta_\mu$.

The bowing to the right of the contours in Fig.~\ref{fig:m12min} is a
result of cancellations between different contributions to the EDMs
\cite{FOS}, and we can see that the effect is to relax the upper bound
on $\theta_\mu$ by a factor of a few.  As we increase $A_0$, the
extent of the bowing increases, and larger values of $\theta_\mu$ can
be accessed.  This loophole to larger $\theta_\mu$ is limited by the
diminishing size of the regions in which there are sufficient
cancellations to satisfy the EDM constraints.  In general, the regions
of cancellation for the electron EDM are different than those for the
${}^{199}$Hg EDM, and the two regions do not always overlap.  As
$\theta_\mu$ is increased, the sizes of the regions of sufficient
cancellations decrease; in Fig.~\ref{fig:m12min}, the width in $\m12$
of the combined allowed region near the $\theta_\mu$ upper bound is
40-80 GeV, which on a scale of 200-300 GeV is reasonably broad.
Larger $A_0$ permits larger $\theta_\mu$, but the region of
cancellations shrinks so that a careful adjustment of $\m12$ becomes
required to access the largest $\theta_\mu$.  At the end of the day,
values of $\theta_\mu$ much greater than about $\pi/10$ cannot satisfy
the EDM constraints without significant fine-tuning of the mass
parameters.  At larger values of $\tb$, the upper bound decreases
roughly as $1/\tb$.  See \cite{fopr} for more details on the status of
EDM and cosmological constraints on CP violating phases in mSUGRA.

\section{Summary}

In summary, constraints on the relic abundance of LSP neutralinos
place significant restrictions on the parameter space of the MSSM,
and mSUGRA in particular.  To accurately compute the cosmological
upper limits on MSSM masses requires the inclusion of coannihilation
effects, both for the case of a Higgsino and gaugino like neutralino.
In particular, we have found that slepton-neutralino coannihilation greatly affects the
neutralino relic abundance when the neutralino and slepton are closely 
degenerate in mass, as is the case in mSUGRA near where the cosmological upper
bound on the neutralino mass is saturated.  Including coannihilation
effects significantly relaxes the cosmological bounds on $\m12, m_0$
and $\mchi$, and reduces the combined Higgs + cosmology lower bound on 
$\tb$, although the upper bounds on CP violating phases in
mSUGRA are not relaxed.  The reach of the LHC will be needed to be sensitive
the full cosmologically allowed region.   Lastly, although I did not discuss
it here, similar effects are present for a gaugino like neutralino in
the general MSSM.   

\vspace{0.5in}
\noindent{ {\bf Acknowledgments} } \\
\noindent  {The work of T.F. was supported in part by DOE   
grant DE--FG02--95ER--40896 and in part by the University of Wisconsin  
Research Committee with funds granted by the Wisconsin Alumni Research  
Foundation.}

\end{document}